\documentclass[fleqn,usenatbib]{mnras}

\usepackage{newtxtext,newtxmath}

\usepackage[T1]{fontenc}
\usepackage{ae,aecompl}

\usepackage{graphicx}   
\usepackage{amsmath}    
\usepackage{amssymb}    

\usepackage{color}

\newcommand\be{\begin{equation}}
\newcommand\en{\end{equation}}
\newcommand\msun{M_{\odot}}

\newcommand\mdot{\dot{M}}
\newcommand\msunyr{M_{\odot}\, {\rm yr^{-1}}}

\newcommand\gcmtwo{\rm g\, cm^{-2}}

\title[T Tauri accretion ] {How do T Tauri stars accrete?}
  \author[Hartmann \& Bae]
      {Lee Hartmann$^{1}$, Jaehan Bae$^{1}$ \thanks{E-mail:
          lhartm@umich.edu} 
\\
$^{1}$Department of Astronomy, University of Michigan,  
	1085 S. University Ave, Ann Arbor, MI 48105, USA \\
}

\date{Accepted XXX. Received YYY; in original form ZZZ}

\pubyear{2017}

\begin{document}
\label{firstpage}
\pagerange{\pageref{firstpage}--\pageref{lastpage}} \maketitle

\begin{abstract}
We conjecture that observed protoplanetary
disc accretion rates may be explained with
low viscosities which could be the result
of hydrodynamic turbulence. We show that
viscosities parameterized in the usual way
with $\alpha \gtrsim 10^{-4}$, comparable to values suggested for hydrodynamic turbulence, can
explain the observed accretion rates and
lifetimes with plausible inner disc surface densities.  Our models are also in better
agreement with surface density estimates
of the minimum mass solar nebula than models
with rapid transport for a given mass accretion
rate, such as recent models of accretion driven
by magnetic winds. 
The required surface densities are a natural
result of the protostellar infall phase, as long
as non-gravitational transport is limited.
We argue that, in addition to possible
non-ideal magnetic transport due to disc winds
possibly modified by the Hall effect, the
effects of low-viscosity hydrodynamic accretion deserve more consideration.

\end{abstract}


\section{Introduction}

T Tauri stars accrete significant amounts of gas over typical timescales of a few Myr, but it has proved difficult to identify precisely why they do so. It was recognized early on that the magnetorotational instability (MRI), which provides a satisfactory explanation for turbulent transport in ionized astrophysical discs \citep{balbus98}, was unlikely to be effective
in large regions of cold protoplanetary discs.  To explain the observed mass accretion in T Tauri
stars, \cite{gammie96} proposed that cosmic rays could sufficiently ionize upper disc layers for the MRI to operate, leaving a non-turbulent, non-accreting ``dead zone'' sandwiched in between. However, Gammie's model included only Ohmic dissipation; recent numerical simulations indicate that ambipolar diffusion quite effectively quenches the MRI over large regions of discs \citep[e.g.,][]{baistone13,bai14,gressel15,simon15}, and
any reduction of cosmic ray ionizing fluxes by disc winds would make MRI activation even less
likely \cite{cleeves15} \citep[although there may be
some MRI activation in the outermost regions;][]{simon15}.

Observations suggesting low turbulence in protoplanetary discs also limit the possible levels of MRI turbulence. \cite{flaherty15} analyzed line profiles of
differing isotopologues of CO to place low limits on turbulent velocities in the outer disc of HD 163296. 
More indirectly, low turbulence is needed for models in which dust concentration in pressure maxima explain observations of narrow rings, gaps,
and possibly vortex structures
\citep{zhubaruteau16,bae17,dong17}.
Finally, several recent studies have
found no evidence for the correlations
between disk masses, sizes, and accretion
rates expected for strongly-viscous disks
\citep{rafikov17,tazzari17,lodato17}.

The time-dependent non-ideal MHD disc simulations which demonstrated the ineffectiveness of the MRI in regions
where ambipolar diffusion is important
instead found strong disc winds which drive accretion in narrow upper disc layers \citep{bai14,bai15,gressel15},
echoing the pioneering study of \cite{konigl89}.
\cite{bai16} has developed one of the most complete pictures of how winds can drive
the necessary accretion in protoplanetary discs,
updating earlier ideas 
\cite{pudritz83,ferreira93,ferreira06}.
The Hall effect in protoplanetary discs may also
modify this picture \citep[e.g.,][]{salmeron08},
as it is capable of producing strong
laminar stresses if the magnetic field 
and angular momentum vector are aligned, or intermittent turbulent stresses in the 
anti-aligned case \citep{lesur14,simon15,bethune17,baistone17}.

However, the effectiveness of non-ideal magnetic
transport depends on uncertain or unknown
parameters, such as the exact levels of ionization
in upper disc layers and magnetic field strengths
\citep[e.g.,][]{armitage13}.
The amount of magnetic flux  
dragged in from the protostellar cloud
and retained in the disc must be
regarded as uncertain 
\citep{guilet14,okuzumi14,baistone17}, 
especially as there are no direct
in situ constraints, and because magnetic flux 
loss via ambipolar diffusion as the protostellar
envelope collapses is likely to be very important, if not essential, to forming 
discs of the observed $\sim 100$~AU sizes \citep{masson16}.  

Observational evidence in favor of the
magnetic models is limited.  While
the jets and winds from inner discs
almost certainly require magnetic acceleration
\citep[e.g.,][]{ferreira06}, these may actually
arise from warmer regions where 
the MRI can operate.  There is evidence for
slower outer disc winds in the form of
low-velocity forbidden line emission  \citep{pascucci11,rigliaco13}, but it isn't clear
whether these flows might be entirely the
result of photoevaporatively-driven mass loss with mininal or no magnetic coupling \citep{msimon16}.
Finally, the bimodality
of the Hall effect depending upon the alignment of poloidal field and angular momentum,
or other distinct differences such as
quenching
accretion, one-sided winds, etc.\ seen in
the simulations of \cite{bethune17}  
currently do not have obvious observational
support \citep[see also][]{baistone17}.

In view of
the likely weak coupling between magnetic fields
(if any) and gas in much of protoplanetary discs, increased attention has been paid to hydrodynamic sources of turbulence
\citep[see][for a review]{fromang17}, such as the vertical shear instability \citep[VSI;][]{urpin03,nelson13,lin15}
convective overstability \citep{klahr14},
the ``zombie vortex'' instability (\citealt{marcus15}; but see \citealt{lesur16}),
or even turbulence driven by planets 
\citep{bae16,fung17}.  
While these instabilities can in principle produce angular momentum transport, the levels of viscosity produced are so low that accretion during T Tauri lifetimes could only come from inner regions of discs; viscous transport timescales at large radii
are too long
compared with observed disc lifetimes.  Conventional disc mass estimates suggest that
the needed reservoir to maintain disc accretion
over typical lifetimes must extend to the outer
disc (\S 2).  However, disc masses 
derived from mm- and sub-mm wave dust emission
are sensitive to uncertain dust opacities,
and inner disc dust is generally thought to be optically thick at these wavelengths (\S 4), thus providing no observational constraints.

In this paper we explore the possibility that protoplanetary disc (T Tauri) accretion can be explained with low turbulent viscosities potentially achievable by a variety of hydrodynamic instabilities.  \cite{fung17} made the suggestion that turbulence driven by a series of super-Earths in inviscid discs could explain the observed accretion rates of $\gtrsim 10^{-8} \msunyr$; here we consider the more general situation of weak viscous transport, regardless
of its origin.  We show that plausible
inner disc masses can provide the needed 
reservoirs of material, such that the 
observed accretion can be explained by a 
combination of low-viscosity, potentially purely hydrodynamic,
turbulence and thermal MRI activation in the warm, innermost disc.

\section{Mass reservoirs needed for accretion}
\label{sec:massreservoirs}

The durations and rates of accretion in T Tauri stars place statistical constraints on the minimum amount of gas initially present in their discs. The most reliable measures of mass accretion rates come from excess optical emission produced as gas funnels in along magnetospheric flux tubes and shocks at the stellar photosphere \citep[e.g.,][]{hartmann06}.
 A recent analysis of accretion rate estimates by G. Herczeg \citep[presented in][]{hartmann16}
found a mean dependence of the mass accretion rate on stellar mass of 
\begin{equation}
\mdot = 10^{-7.9} M_*^{2.1}\, \msunyr\,,
\label{eq:mdotmass}
\end{equation}
where the stellar mass $M_*$ is measured in
solar masses. Removing this mass dependence for
stars with masses between $0.3$ and $1.0 \msun$ 
yields a time dependence $\mdot \propto t^{-1.07}$.
Given the large observed
scatter about these trends, we simplify Herczeg's results in this
paper to
\begin{equation}
\mdot = 1.8 \times 10^{-8} M_{0.7}^2 t_6^{-1} \msunyr\,,
\label{eq:mdotmean}
\end{equation}
where $t_6$ is the age in units of Myr.
The total mass accreted between times $t_0$ and $t_f$ is then
\begin{equation}
M_{acc} = 1.8 \times 10^{-2} M_{0.7}^2 \ln(t_f/t_0) \msun\,.
\end{equation}

As this relation was derived for optically-visible young stars, we take $t_0$ to be the time when 
infall to the disc has ceased (i.e., after
the protostellar phase).  \cite{kenyon95} suggested 
that the protostellar phase lasts between 0.1-0.2 
Myr, while the more recent and complete study by 
\cite{evans09} estimated protostellar lifetimes $\sim 0.5$~Myr. 
Here we adopt a compromise value of $t_0 \sim 0.4$~Myr, 
consistent with the fits being derived mostly from stars with larger ages.
The fraction of accreting stars
decreases steadily with increasing
age, such that only about 50\% of all stars show significant accretion at ages of 3 Myr, with a much smaller fraction lasting for 10 Myr
\citep{hernandez07}; the e-folding
time for near-infrared excesses or
measurable accretion to disappear is
estimated to be $\sim 2.3-3.0$~Myr
\citep{fedele10}.
We therefore set $t_f = 3$~Myr, and find 
an estimated average {\em accreted} 
mass during the T Tauri phase of
\begin{equation}
<M_{acc}> \approx 0.036 M_{0.7}^2 \msun\,.
\label{eq:totmean}
\end{equation}

To compare with observational estimates
of disc masses from dust emission,
we suppose that a typical T Tauri star has an age of
$\sim 2$~Myr, and adopt an e-folding
time for the end of accretion
$= 3$~Myr as above.  Then the typical
accretion rate for the fiducial stellar mass of
$M_* = 0.7 \msun$
is $\sim 0.9 \times 10^{-8} \msunyr$
and the mass that will be
accreted later is $\sim 0.01 \msun$.  While
this is in
reasonable agreement with the median disc mass
$\sim 0.005 \msun$ estimated from mm-wave dust emission of stars in the Taurus molecular cloud 
\citep[e.g.,][]{williams11}, it must be emphasized
that this is the accreted mass, and
the disc masses must be larger if there is
to be anything
left to form planets. 

Although the scatter
around the mean values for accretion and
mass reservoirs given in equations \ref{eq:mdotmean} and \ref{eq:totmean} is large, these results are 
important in that infall onto the star
traces the bulk gas,
in contrast to estimates
from dust emission which are sensitive to
the size distributions of the solids and which
do not trace mass in optically-thick regions.
\cite{williams14} attempted to determine gas
masses by analyzing the emission from 
CO isotopologues, but their results yielded
masses an order of magnitude too small to
account for the reservoir of mass needed to
sustain accretion, and thus are probably 
strongly biased by CO freeze-out onto
grains or incorporation into other bodies.

\section{Rapid Mass transport from the outer disc?}
\label{sec:outerdisc}

Because the typical mass reservoirs needed for T Tauri accretion are comparable to (or exceed) usual estimates for entire discs, it is usually assumed that mass must be accreted from large radii during T Tauri lifetimes.  To fix ideas, we consider a disc
around a central star of mass of $0.7 \msun$ and
assume a disc temperature distribution 
$T = 200 (R/AU)^{-1/2}$~K, motivated by the radiative transfer models of \cite{dalessio01}.  
We further assume that the disc
has a viscosity 
\begin{equation}
\nu = \alpha c_s^2/\Omega\,,
\end{equation}
where $c_s$ is the sound speed and $\Omega$ is the Keplerian angular velocity.
Then the viscous timescale as a 
function of radius is then
\begin{equation}
t_{\nu} \sim {R^2 \over \nu} = 
1.2 \times 10^4 \alpha_{-2}^{-1} T_{200}^{-1} M_{0.7}^{-1/2}
{R \over {\rm AU}} \, {\rm yr}\,,
\label{eq:tvisc}
\end{equation}
where $\alpha_{-2} = \alpha/0.01$ and
$T_{200}$ is the disc temperature at 1 AU
in units of 200~K.

Thus, if the mass reservoir for accretion
must extend to $\sim 100$~AU during a T Tauri lifetime of a few Myr, the viscous parameter
must be in the range
$\alpha \sim 10^{-2} - 10^{-3}$
\citep{hartmann98}. While such viscosities could plausibly achieved in an MRI-active disc, this is
unlikely to be the case for protoplanetary discs
(\S I). Moreover,
there is weak or little empirical evidence for the relations between disc size, mass,
and accretion rate predicted by highly viscous disc models \citep{rafikov17,tazzari17,lodato17}.

It is worth noting that high viscosities
with typical T Tauri accretion rates are
incompatible with estimates of mass surface densities for the lowest mass estimates
for the solar nebula \citep[the ``minimum-minimum 
mass solar nebula'' (M-MMSN),][]{weidenschilling77,hayashi81}
let alone larger values for the MMSN
\citep{desch07}, or estimates
of the typical "minimum mass extrasolar nebula"
derived assuming in-situ formation
of super-Earth systems \cite{chiang13}.
Using the steady state result
$\mdot = 3 \pi \nu \Sigma$ to estimate the surface
density
at 1 AU given the accretion rate and viscosity
parameter,
\begin{equation}
\Sigma_1  \sim 200 \mdot_{-8} \alpha_{-2}^{-1} T_{200}^{-1}\, \gcmtwo\,,
\end{equation}
where $\mdot_{-8}$ is the mass loss rate
in units of $10^{-8} \msunyr$. 
Thus, to achieve even the M-MMSN surface density
of  $1700 \, \gcmtwo$ implies $\alpha \sim 10^{-3}$ 
if discs are viscous. While the solar nebula is
obviously only one object, the existence of
many compact extrasolar planetary systems
suggests there may be a need for higher inner disc mass
surface densities difficult to achieve
with high viscosities.

The relationships between accretion rates and
disc surface densities of non-ideal MHD transport models discussed in \S 1 
are not clear, as they are not describable by
a turbulent $\alpha$ viscosity; there may be
some relation due to the dependence of the
stresses on the plasma $\beta$.  Rapid transport
can be consistent in some cases with inner
disc surface densities below the M-MMSN
\citep[see, for example, Figure 6 of][]{bai16}.

\section{Slow Mass transport from the inner disc?}
\label{sec:innerdisc}

Given that mass reservoirs estimated from
dust emission are at best barely sufficient
to sustain typical accretion,
it may seem counter-intuitive
to consider whether the inner disc alone can
supply the needed mass.
However, there are very few observational
constraints on inner disc surface densities,
both because these regions are still mostly
unresolved spatially and the dust emission from the inner disc is expected to be optically thick \citep[see, for example,][]
{tazzari16}.\footnote{Transitional
discs (e.g., discs with large inner holes or gaps 
in dust) have much greater small dust depletions \citep{espaillat14}, 
but these are clearly quite evolved
from their initial conditions.}

To see more quantitatively what is required
for adequate accretion at low viscosities,
we develop a vertically-integrated disc model with both viscous and irradiation heating.  We assume
steady accretion in the inner regions, such that
the surface density is given by
\begin{equation}
\mdot = 3 \pi \alpha c_s^2 \Sigma \Omega^{-1}\,.
\label{eq:mdotvisc}
\end{equation}
The sound speed appropriate to
the central temperature 
\begin{equation}
T_c^4 = T_v^4 + T_i^4\,,
\label{eq:tir_tv}
\end{equation}
where 
\begin{equation}
T_v^4 = {3 \tau \over 8} \,  
{3 G M_* \mdot \over 8 \pi \sigma R^3}\,,
\label{eq:tv}
\end{equation}
where $\tau$ is the vertical optical depth from the midplane \citep{hubeny90}.

We use the low temperature dust opacity from
\cite{zhu09} of $\kappa = 5.3 \times 10^{-2} f T^{0.738} \, {\rm cm^2 g^{-1}}$ when $T < 1500 K$, modified by the factor $f$ to account
for reductions in opacity due to depletion
of small dust which dominates the radiative
trapping in the warmer inner regions of
the disc. Observations, which provide clear
evidence for grain growth in outer
discs \citep[e.g.,][]{tazzari16}, 
implies a reduction of the abundance of
small grains.  More directly, detailed
modeling of spectral energy distributions
require depletions of small dust in
upper disc layers in the range of $10^{-1}$ to $10^{-3}$ \citep{furlan06}. 
We therefore adopt
values of $f = 0.1, 0.01$
as observationally justifiable.

We do not model regions where the above equations
would yield $T > 1500$~K;
we simply assume
the MRI is thermally activated, with a much larger $\alpha$, and thus adjusts the surface
density and temperature to accommodate the
mass accreting from the adjacent colder region.

For a given accretion rate, low viscosities imply
large surface densities, which can approach
values such that transport by gravitational
instability (GI) must be considered.
Numerical simulations indicate 
GI-driven spiral waves readjust disc surface
densities rapidly such that the Toomre parameter 
$Q = c_s \Omega/(\pi G \Sigma)$ approaches
$\approx 1.4$ \citep{boley06}.  At somewhat larger values of $Q$, some transport can still occur with
decreasing efficiency, becoming
negligible at $Q \ge 2$ \citep{kratter08,griv06}.
We therefore limit the surface density such that
when equation \ref{eq:mdotvisc} would imply strong
GI transport, we determine the surface density
assuming $Q = 2$.  This means that the accretion
rate in these regions is no longer constant.

\begin{figure}
\includegraphics[width = \columnwidth]{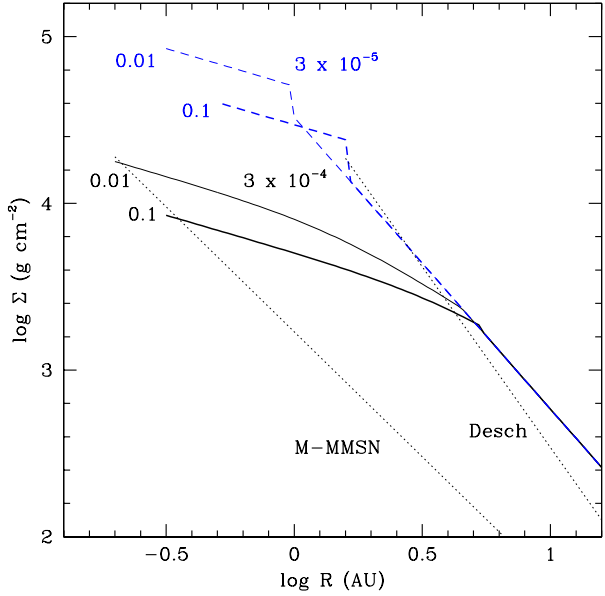}
\caption{Surface density distributions for discs heated by both viscous dissipation and stellar irradiation, for an inner disc accretion rate of $2 \times 10^{-8} \msunyr$. The solid curves are the results for $\alpha = 3 \times 10^{-4}$, with $f = 0.1$ (heavy curve) and $f = 0.01$ (lighter curve). The dashed curves show the surface densities for $\alpha = 3 \times 10^{-5}$, with $f = 0.1$ and 0.01 for the heavy and lighter lines, respectively. The transition to a steeper decrease in $\Sigma$ at larger radii indicates the regions where the $Q = 2$ GI-limit comes into effect (see text). The jump in surface density is due to the simple way we switch between steady accretion and the GI-limit. The outer dotted line shows the $\Sigma$ for the MMSN of Desch (2007) while the inner line is the M-MMSN.}
\label{fig:sigma}
\end{figure}

Figure \ref{fig:sigma} shows the 
surface density distributions
for the  fiducial $0.7 \msun$ star 
with  $T_i = 200 (R/AU)^{-1}$~K as before, 
for accretion rates of
$\mdot = 2 \times 10^{-8} \msunyr$,
$\alpha = 3 \times 10^{-4}, 3 \times 10^{-5}$,
and $f = 0.1, 0.01$. With the larger
viscosity parameter, the surface density
distribution is consistent
with steady accretion out 
to about 5 AU, while for
$\alpha = 3 \times 10^{-5}$ there is only
a very narrow region interior to $\sim 1-2$ AU that is
in steady state, depending upon $f$.
In general, lower opacities lead to somewhat higher surface densities as the amount of trapping of viscously-generated heat is reduced.  In the regions where $Q=2$ limits
the surface density, the accretion rate is
no longer constant but decreases
with increasing radius. This means that
once the inner regions drain, and accretion
is sustained from the GI-limited regions,
the accretion rate into the central regions
should decrease, as we verify in the following
section with time-dependent models.
The surface densities are well above the M-MMSN but are comfortably below the MMSN of \cite{desch07}
for $\alpha = 3 \times 10^{-4}$ and even 
consistent for $\alpha = 3 \times 10^{-5}$. 


\begin{figure}
\includegraphics[width=\columnwidth]{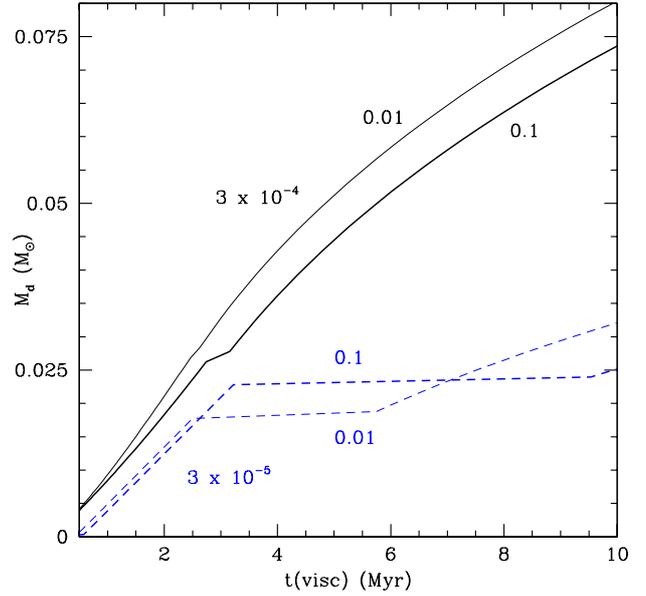}
\caption{The enclosed mass as a function of the viscous timescale $R^2/\nu$
for the model surface densities shown in Figure \ref{fig:sigma}, with the
same line types for the different cases.}
\label{fig:masstv}
\end{figure}

In Figure \ref{fig:masstv} we plot the mass interior to the radius $R$ where the viscous timescale is
$t_{\nu}$.  Both high mass loss rate models 
show approximately the same mass, $\sim 0.02 \msun$, within
a radius where $t_{\nu} \sim 3$~Myr; although
the steady region is smaller at low viscosity,
the surface density is higher and the viscous
timescales are longer at a given radius.  
The transition to the $Q=2$ region 
is evident 
from the breaks in the curves. 
The model mass reservoirs are
within a factor of two of the observational
estimate of equation \ref{eq:totmean}, and
thus these simple calculations suggest
that the necessary accretion rates can
be maintained over 
timescales $\sim 3$~Myr.

\begin{figure*}
\includegraphics[scale=0.6]{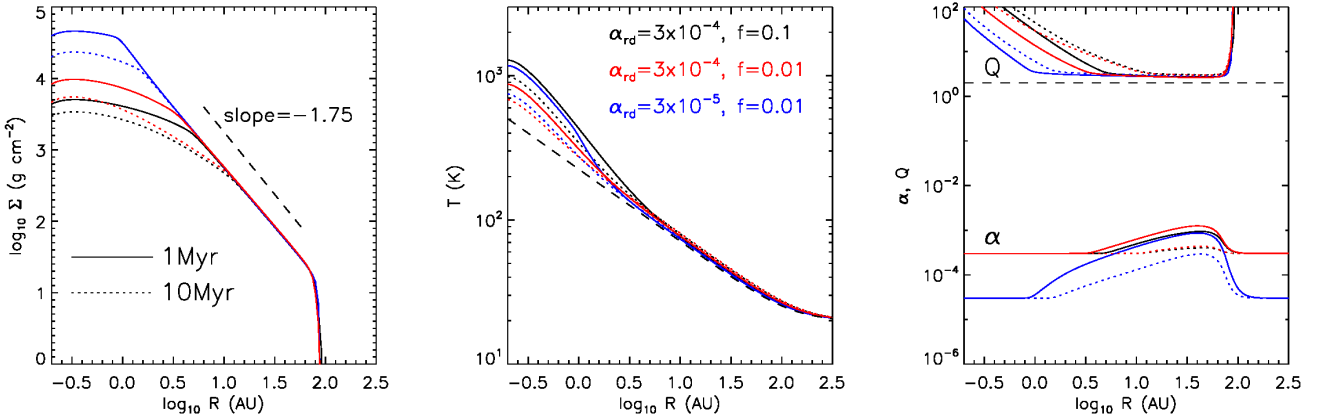}
\caption{The radial distributions of (left) the surface density, 
(middle) the temperature, and (right) viscosity parameter $\alpha$ and Toomre $Q$ parameter. 
In each panel, black curves present results with $\alpha_{\rm rd}=3\times10^{-4}$ and $f=0.1$, red curves present results with $\alpha_{\rm rd}=3\times10^{-4}$ and $f=0.01$, and blue curves present results with $\alpha_{\rm rd}=3\times10^{-5}$ and $f=0.01$, respectively. 
The solid and dotted curves are radial profiles at 1 and 10~Myr, respectively. In the left panel, the dashed curve indicates a power-law slope of $-1.75$.
In the middle panel, the dashed curve indicates the external irradiation-dominated temperature for $M_* = 0.7~\msun$; the excess temperature above this line is due to viscous heating. 
In the right panel, the dashed curve indicates $Q=2$.}
\label{fig:disc_profile}
\end{figure*}

\section{Low-viscosity disc evolution with infall}
\label{sec:timedep}

Whether T Tauri discs have relatively large
inner disc surface densities as shown
in Figure \ref{fig:sigma} depends on initial conditions
as well as requiring low viscosities (or other, slow,
transport).  To explore the possible initial
disc mass distribution we employ a simplified
version of the methods used in \cite{bae13} and
\cite{zhu10} to follow disc evolution with infall.
As in those previous investigations, we use
a model of infall to the disc motivated by
the \cite{terebey84} model for the collapse
of a rotating protostellar cloud, but these
one-dimensional calculations do not include
an active layer.
Here, we briefly summarize the time-dependent model with infall; for more details we refer readers to Section 2 of \cite{bae13}.

We solve the one-dimensional mass and angular momentum conservation equations to evolve the disc surface density,
adopting the infall model of \cite{cassen81} modified as in \cite{bae13}.
The two equations combine into a diffusion equation as
\begin{eqnarray}
\label{eqn:diffusion}
\dot{M} & = & 6\pi R^{1/2} {\partial \over \partial R} (R^{1/2} \Sigma \nu) + {2\pi R^2 \Sigma \over M_R} {\partial M_R \over \partial t} \nonumber \\
& & - 4\pi \left( {R \over GM_R}\right)^{1/2} (\Lambda(R,t)-g(R, t) R^2 \Omega(R)),
\end{eqnarray}
where $M_R$ is the sum of the mass of the central star and the disc within a radius $R$ and $\Lambda$ and $g$ describe the angular momentum and mass flux per unit distance from the infalling material.
In the diffusion equation, the first term represents disc accretion due to viscosity, the second term represents the mass redistribution due to the change in central stellar mass over time, and the third term represents the accretion arising due to the infalling material.

The disc temperature is computed by balancing heating and radiative cooling,
\begin{equation}
C_\Sigma {\partial T \over \partial t} = Q_{\rm heat} - Q_{\rm cool},
\end{equation}
where $C_\Sigma = \Sigma c_s^2 / T$ is the heat capacity.
The heating term $Q_{\rm heat}$ consists of the viscous heating, the heat generated by the shock dissipation of infalling material, and the external irradiation which includes the stellar luminosity, accretion luminosity, and the heat from the background envelope cloud ($T_{\rm bg}=20$~K).
The radiative cooling rate $Q_{\rm cool}$ is calculated as 
\begin{equation}
Q_{\rm cool} = {16 \over 3} \sigma T^4 {\tau \over 1+\tau^2},
\end{equation}
where $\tau=f \kappa \Sigma/2$ is the optical depth of the disc and $f \kappa $ is
the Rosseland mean opacity used in the
previous section for the quasi-steady models.

We implement a total viscosity parameter $\alpha= \alpha_{\rm rd} + \alpha_{\rm GI} + \alpha_{\rm MRI}$, where $\alpha_{\rm rd}$ accounts for any non-GI/MRI transport, which we assume constant at a level $3\times10^{-4}$ or $3\times10^{-5}$, $\alpha_{\rm GI} = e^{-Q^2}$ accounts for the transport associated with GI, and $\alpha_{\rm MRI}$ accounts for the transport associated with the MRI, which is set to 0.01 only when the disc temperature exceeds the MRI-activation temperature $T_{\rm MRI}=1500$~K.

The calculations start with a $0.1~M_{\odot}$ central protostar surrounded by an $1~M_{\odot}$ cloud.
We assume a rigid rotation for the cloud, with a $3~\%$ of the breakup angular frequency at the outer cloud edge. 
Infalling material is added to the disc at a constant rate $\sim3.4\times10^{-6}~\msunyr$ for
$\sim 0.24$~Myr, forming a $M_* \sim 0.65 \msun$ star with a $\sim0.25 \msun$ of surrounding disc at the end of the infall phase.
In this model, most of the stellar mass is accreted during the infall phase through outbursts \citep[see, e.g.][]{bae13,zhu10}.
The centrifugal radius, marking the outer edge of the disc where matter is
infalling, moves outward with time,
reaching $\sim 25$~AU at the end of the infall phase.  

\begin{figure*}
\centering
\includegraphics[scale=1]{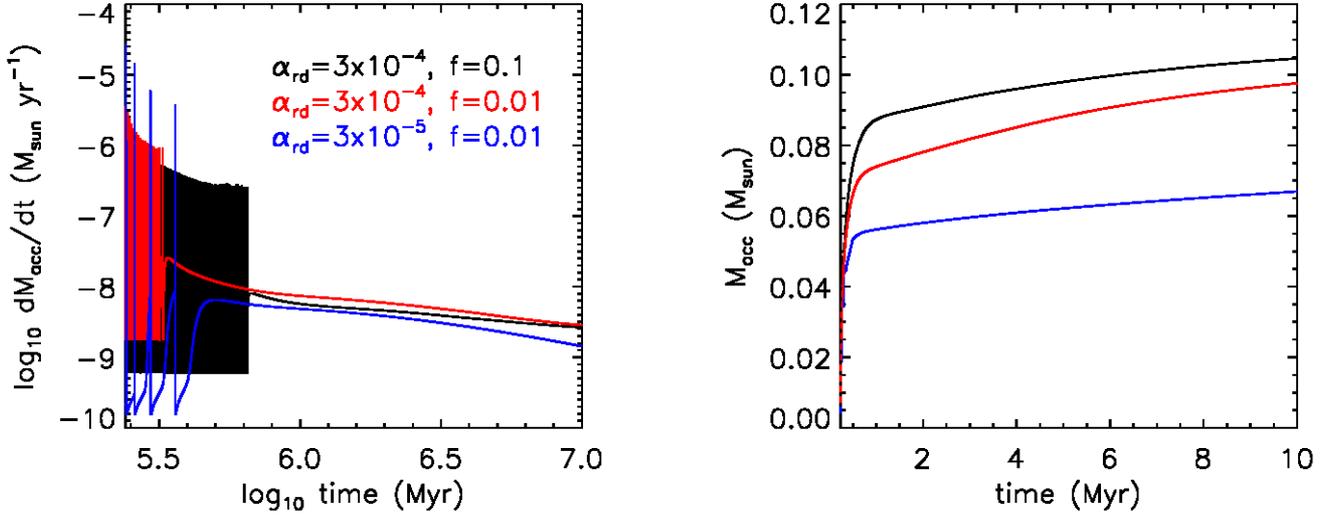}
\caption{(Left) The accretion rate and (right) total mass accreted as a function of time for (black) $\alpha_{\rm rd}=3\times10^{-4}$ and $f=0.1$, (red) $\alpha_{\rm rd}=3\times10^{-4}$ and $f=0.01$, and (blue) $\alpha_{\rm rd}=3\times10^{-5}$ and $f=0.01$ models. Note that the accreted mass presented here is during the post-infall, T Tauri phase only ($t > 0.24$~Myr).}
\label{fig:mdot}
\end{figure*}

In Figure \ref{fig:disc_profile} we present the radial distributions of surface density, temperature, $\alpha$, and $Q$ at 1 and 10~Myr for
the fiducial model with $\alpha_{\rm rd} = 3\times10^{-4}$ and $f=0.01$.
The overall surface density profile is in good agreement with that of the quasi-steady
disc model (Figure \ref{fig:sigma}).
The disc expands beyond the final
infall radius of 25 AU 
to $\sim 100$~AU due to the action
of gravitational torques,
which redistribute material such that the disc has a constant $Q\sim3$ and
thus a power-law density profile  $\Sigma \propto R^{-1.75}$ as a result
of the $T \propto R^{-1/2}$ temperature
distribution in the irradiation-dominated regime.
Inward of the $Q$-limited region,
the disc has a roughly constant mass accretion as a function of radius,
as assumed in the quasi-steady model.  The
surface density is flatter because
the temperature is dominated
by viscous heating, enhanced by
radiative trapping due to the large surface density of $\Sigma > 10^3~\gcmtwo$.
Note that, without photoevaporation and planet formation, the outer disc surface density remains nearly constant over 10~Myr because of the long viscous timescale for a small viscosity $\alpha_{\rm rd}$.
It is only the inner disc that maintains the accretion onto the central star while the total disc mass is dominated by the outer disc.  
Thus, the model
predicts no relation between disc size,
mass, and accretion rate, in agreement
with the study of \cite{rafikov17}.

In Figure \ref{fig:mdot} we present the mass accretion rate history, along with the total mass accreted onto the central star during the post-infall (T Tauri) phase ($t > 0.24$~Myr).
The disc enters a quiescent accretion phase following a number of outbursts after the infall ceases.
The accretion rate during the extended quiescent accretion phase starts with $\dot{M}\sim$ a few $\times10^{-8}~\msunyr$ and gradually decreases over time.  While the
magnitude of the accretion rate and
the amount of mass
accreted ($\sim0.02~\msun$ between 1 and 10 Myr) is in reasonable agreement
with median observed values,
the best fit power-law slope for the $\log t - \log \dot{M}$ relation between 1 and 10~Myr is $-0.45$, shallower than
the estimated $\sim-1$ (\S 2).
For this model a faster decrease in accretion rate would have to be explained
by photoevaporation or planet formation.

One effect of our adoption of reduced
dust opacities is that it reduces or
eliminates the accretion outbursts lasting
well through the T Tauri phase seen in
our previous simulations
\citep{zhu10,bae13}, outbursts that
are inconsistent with observations.  
Reducing the opacity reduces or eliminates
the T Tauri phase outbursts by
reducing the amount of radiative
trapping essential to drive temperatures
high enough to trigger the MRI.
This accounts for the difference in
early accretion behavior 
between the $f = 0.1$ and 0.01,
$\alpha_{\rm rd} = 3 \times 10^{-4}$
models (left panel of
Figure \ref{fig:mdot}), such that larger
depletion stops the outbursts
sooner.
Changing $f$ by a factor of 10 
has little effect on the accretion rates
because it has only a modest effect on the surface density
in the viscous accretion (inner) zone, and no effect on the
Q-limited zone where the temperature is dominated by the assumed
irradiation flux (see Figure \ref{fig:sigma}).

With $\alpha_{\rm rd} = 3 \times 10^{-4}$,
the more numerous
accretion outbursts for $f=0.1$ case
reduce the surface density compared
with the results for $f = 0.01$
(Figure \ref{fig:disc_profile}).
However, the mass
accretion rate and mass reservoirs
are still comparable and within
observational requirements.
With a lower level of non-GI/MRI transport ($\alpha_{\rm rd} = 3\times10^{-5}$), the surface density in the inner disc is higher and the $Q$-limited outer disc starts at a smaller radius of only $\sim1-2$~AU,
again similar to that of the quasi-steady
model.
The disc maintains an accretion rate
of a few $\times10^{-9}~\msunyr$ during the T Tauri phase, accreting in total $\sim 0.01~\msun$ of mass onto the central star.
As time goes on, mass is continually drawn from the $Q$-limited disk regions where
$\Sigma \propto R^{-7/4}$; as this is steeper than the $R^{-1}$ dependence for a steady constant $\alpha$ disk with $T \propto R^{-1/2}$
\citep{hartmann98}, the accretion rate $\mdot \propto \nu \Sigma$ must decrease with time.  This
decrease in our models is slow because of the long viscous times; it occurs somewhat faster in the
low-$\alpha$, $f =0.01$ case because the $Q$-limited region becomes the mass reservoir sooner (see Figure \ref{fig:masstv}).

\section{Discussion}

The calculations of the previous section show that in the absence of rapid disc transport, protostellar collapse phase is likely to produce initially massive discs.
Viscous or other transport timescales
must be comparable to or shorter than
typical low-mass protostellar lifetimes
of $\sim 0.4$~Myr 
(\S \ref{sec:massreservoirs}) to avoid piling
up mass in the disc; at scales of 10-100
AU, this requires viscous or equivalent
$\alpha$ parameters $\sim 10^{-2}$ or
larger 
\citep[Equation \ref{eq:tvisc};][] {kratter08,zhu09}. GI transport can
redistribute mass effectively but
by itself would still yield a massive disc. The effectiveness of
wind transport during the protostellar phase is
questionable, as the wind could
be impeded or even quenched by the
higher ram pressures of the infalling
material.  Dust in the infalling
envelope might also extinct the
ionizing stellar radiation needed for
effective coupling of the magnetic
field to the gas, which could also
limit Hall effect transport.  Finally,
observations of protostellar
accretion luminosities do not provide
evidence for rapid accretion during
this phase \citep[see discussion in][]{hartmann16}.

The most direct evidence for large 
inner disc masses during or near the
end of protostellar infall
comes from study of the outbursting FU Ori objects
\citep{hartmann96}. FU Ori itself
has a central gravitating mass (protostar)
of $\sim 0.3 \msun$ accreting
at a rate $\sim 2 \times 10^{-4} \msunyr$.  FU Ori has been in a rapid accretion state for the $\sim 80$~yr duration of its outburst, and thus has accreted $\sim 0.02 \msun$ - about
the same that equation \ref{eq:mdotmass} predicts is accreted on average by a $0.3 \msun$ star
over its entire T Tauri disc lifetime.  Analysis of the spectral energy distribution of FU Ori
indicates that the rapid accretion region extends over $\sim 0.5 - 1$~AU in radius \citep{zhu07}; 
presumably the accreted mass has been drawn from this region.
While FU Ori objects may not be typical
of late stages of protostellar 
evolution, they do suggest that massive
inner discs should be present in at least
some objects at the start of the T Tauri phase.

If 1-Myr-old T Tauri inner discs have low masses, but started out relatively massive at the end of infall, they must
have had high accretion rates and/or photoevaporative rates at
earlier times. The observational
estimates discussed in
\S 2 provide little or no
evidence for sufficiently rapid
early but post-infall
accretion. Photoevaporative mass loss rates are uncertain, but if they
are too large it would be difficult
to explain those T Tauri systems
whose disc lifetimes reach $\sim 10$~Myr.

Because the mass accretion in current
disc wind models occurs only in thin
layers in the upper disc, the magnetic
transport is somewhat decoupled from
the evolution of the central disc regions\citep[but this may be modified by Hall effect:][]{bethune17,
baistone17}.  The \cite{bai16} fiducial
model adopts a residual hydrodynamic
viscosity parameter of $2 \times
10^{-4}$.  As this model has a surface density at 1 AU
about five times smaller than that of
the M-MMSN, the viscously-driven
accretion is
negligible compared with that of the
wind; however, if the
initial values of $\Sigma$ were much
larger, in better agreement with
other solar nebula estimates
\citep{weidenschilling77,desch07},
accretion due to viscous transport would
have been comparable to that of the
wind.

The models discussed here do not explain
either the stellar mass- or time-dependence
of accretion rates. \cite{dullemond06}
attempted to explain the mass dependence
as a result of initial disc sizes with
viscous evolution. Lower-mass stars
tend to have smaller discs (because
they have smaller initial protostellar
cores), and so they viscously evolve
much faster, such that their accretion
rates at a given age are much lower
than those of larger stars, with
larger discs and slower viscous
depletion.  Whether this explanation 
of the dependence of accretion on stellar mass
can work with the much lower viscosities envisaged in 
this paper, or in concert with wind and/or Hall
effect transport, is not clear.

Identifying the factors which determine disk lifetimes
necessarily requires an explanation of the observed disk clearing timescales,
which are much shorter than those expected from viscous evolution
\citep[the 'two-timescale' problem;][]{clarke01}.
Models with photoevaporatively-driven winds \citep[e.g.,][]{clarke01,owen11} 
can in principle explain the two-timescale behavior; the photoevaporative
mass loss opens a gap in the disk, preventing replenishment of the inner disk gas 
that then drains onto the central star.  
\citep[A giant planet could also open a gap to accomplish the same thing;][]{perezbecker11a,perezbecker11b}.) 
However, with the low viscosities envisaged here, the inner disk clearing timescale would be too
long.  Possibly a disk wind working in concert with photoevaporation or planet gap formation 
could be a viable explanation.  Alternatively, turbulence generated by planets in the inner disk 
\citep{fung17} or by an outer giant planet \citep{bae16} might be able to enhance transport sufficiently
to clear the inner disk.


 
\section{Summary}

We have shown that 
viscous disc models with $\alpha \gtrsim 10^{-4}$ can explain 
observed T Tauri mass accretion rates
and lifetimes provided that mass surface densities are sufficiently large.  The required values of 
$\Sigma$ are not in conflict with
any observational constraints nor
do they imply gravitationally unstable discs, and they are comparable to more recent estimates of solar and extrasolar nebula surface densities.
The low viscosities are also consistent with
observations limiting turbulence, providing favorable conditions for dust growth. Our considerations do not rule out the possible dominance of magnetic transport, but show that magnetic stresses may not be essential if hydrodynamic instabilities occur at
low levels.

We acknowledge a helpful report from an anonymous referee.
This work was supported in part
by NASA grant NNX17AE31G and used computational resources and services provided by Advanced Research Computing at the University of Michigan, Ann Arbor.

\bibliographystyle{mnras}
\bibliography{refs1}

\label{lastpage}

\end{document}